\documentclass[fleqn,usenatbib]{mnras}

\usepackage{newtxtext,newtxmath}

\usepackage[T1]{fontenc}

\DeclareRobustCommand{\VAN}[3]{#2}
\let\VANthebibliography\thebibliography
\def\thebibliography{\DeclareRobustCommand{\VAN}[3]{##3}\VANthebibliography}

\usepackage{graphicx}	
\usepackage{amsmath}	
\newcommand{\teff}{$T_\mathrm{eff}$}
\usepackage[dvipsnames]{xcolor} 

\title[New hot subdwarf binaries]{Identification of new hot subdwarf binary systems by means of Virtual Observatory tools}

\author[Solano et al.]{
E. Solano$^{1,2}$\thanks{E-mail: esm@cab.inta-csic.es (ES)},
A. Ulla$^{3}$,
E. P{\'e}rez-Fern{\'a}ndez$^{3,4}$,
C. Rodrigo$^{1,2}$,
R. Oreiro$^{3,5}$,
A. Aller$^{1,2}$,\newauthor
M. Manteiga$^{6}$,
R. Santove\~{n}a-G{\'o}mez$^{7}$,
M. A. {\'A}lvarez$^{7}$,
C. Dafonte$^{7}$
\\
$^{1}$Departmento de Astrofísica, Centro de Astrobiología (CSIC-INTA), ESAC Campus, Camino Bajo del Castillo s/n,
E-28692 Villanueva de la Cañada, Madrid, Spain\\
$^{2}$Spanish Virtual Observatory, Spain\\
$^{3}$Universidade de Vigo (UVIGO), Applied Physics Department, Campus Lagoas-Marcosende, s/n, 36310 Vigo, Spain\\
$^{4}$IES de Beade, Consellería de Educación e Ordenación Universitaria, Camino do Outeiro 10, 36312 Vigo, Spain\\
$^{5}$IES Ramón Llull, Carrer de Ramon Llull, 10, 46021 València, Spain\\
$^{6}$CIGUS CITIC - Department of Nautical Sciences and Marine Engineering, University of A Coru\~na, Paseo de Ronda 51, A Coru\~na, 15011, Spain\\
$^{7}$CIGUS CITIC - Department of Computer Science and Information Technologies, University of A Coru\~na, Campus Elviña s/n, A Coru\~na, 15071, Spain\\
Coru\~na, Spain\\
}

\date{Accepted 2022 June 02. Received 2022 May 24; in original form 2022 March 30}

\pubyear{2021}

\begin{document}
\label{firstpage}
\pagerange{\pageref{firstpage}--\pageref{lastpage}}
\maketitle

\begin{abstract}
The estimation of the binary fraction of hot subdwarfs is key to shed light on the  different evolution scenarios proposed to explain the loss of the hydrogen envelope during the red giant branch phase. In this paper we analyse the spectral energy distribution of the hot subdwarfs included in a recent and  comprehensive catalogue with the aim of identifying companions. Our methodology shows a performance superior to the photometric criteria used in that study, identifying 202 objects wrongly classified as binaries according to their spectral energy distributions, and finding 269 new binaries. Out of an initial sample of 3\,186 objects, we classified 2\,469 as single and 615 as binary hot subdwarfs. 
The rest of the objects (102) were not classified because of their 
inadequate
spectral energy distribution fitting due, in turn, to poor quality photometry.
Effective temperatures, luminosities and radii were computed for 192 singles and 42 binaries. They, in particular the binary sample, constitute an excellent dataset to further perform a more careful spectroscopic analysis that could provide detailed values for the chemical composition, masses, ages, rotation properties or reflection effects for the shortest-period systems. 
The results obtained in this paper will be used as a reference for a forthcoming work where we aim to generalize binary and single hot subdwarf classification using Artificial Intelligence-based techniques.


\end{abstract}




\section{Introduction}
\label{intro}

%
 
Hot subdwarf (hot sd) stars are core-helium burning stars at the
blue end of the horizontal branch or even beyond that stage, in the region known as the extreme horizontal branch. They are often located at high galactic latitudes and, according to atmospheric composition, they are mainly divided into B- (sdB) or O- (sdO) types, if hydrogen or helium dominated, respectively \citep{Drilling13}.

With
effective temperatures exceeding 19 000\,K, logg $\geq$ 4.5 dex, radii of a few tenths of a solar radius and masses of $\sim$ 0.5 M$_{\odot}$, hot sds
are objects that have lost most of their hydrogen envelope (M$_{env}$ $<$ 0.01 M$_{\odot}$) during the red giant branch phase \citep{Heber09}. Therefore, they are unable
to sustain hydrogen shell burning and thus, to follow canonical evolution through the asymptotic giant branch proceeding, instead, directly towards the white dwarf cooling track. Circumstances that lead to the removal of all but a tiny fraction of the hydrogen envelope are still a matter of debate. Theoretical evolution scenarios proposed so far include enhancement of the mass-loss efficiency near the red giant branch tip \citep{Cruz96} or mass loss, mass transfer or colaescence through binary interaction \citep{Pelisoli20}. Further and comprehensive information on the hot sd properties can be found in \citet{Heber16}.

Since the first large area surveys made in the second half of the 20th century \citep[e.g.][]{Kilkenny88}, the number of known hot sds has dramatically grown. In part, this is thanks to the advent of large spectroscopic surveys like MUCHFUSS \citep{Geier11}, SDSS \citep{Geier15, Kepler15, Kepler16}, or LAMOST \citep{Lei19}. \citet{Geier20} includes a recent and up to date compilation of hot subdwarf catalogues. 

In parallel to this data increase, the advent of initiatives like the Virtual Observatory (VO)\footnote{http://www.ivoa.net}  has opened new lines of work in the field of massive data analysis. \citet{Oreiro2011}, using a VO methodology, defined a procedure particularly designed to uncover uncatalogued
hot sds among blue object samples, with a low contamination factor over white dwarfs, cataclysmic variables and OB stars. Out of 38 new candidates, 30 of them with spectra available, they could confirm 26 new hot sds and reported, at least, 8 binary systems. The methodology described in that paper was further extended by \citet{Perez16} who reported 65 new hot sds out of 68 candidates with SDSS spectrum. 

It is well known that a significant fraction of hot sds resides in binary systems. Studies based on optical colours, radial velocities, high-resolution optical spectroscopy or excess in the near-infrared have led to a fraction of hot subdwarf binaries ranging from 20\% to 66\% depending on the methodology, inhomogeneity of data sets or selection effects \citep{Girven12}. 

In this paper we will test the ability of VO tools to uncover new hot sds in binary pairs, and/or to confirm proposed binary candidates using the \citet{Geier20} catalogue of known hot sd stars. The article is organised as follows: after the Introduction (Sect.\,\ref{intro}), we describe in Sect.\,\ref{analysis} the object selection, which is based on good determinations of $Gaia$ EDR3 parallaxes and colours and on the expected locus of hot sds in a colur-absolute magnitude diagram (Sect.\,\ref{sel}),  the procedure followed to build the Spectral Energy Distributions (SEDs, Sect.\,\ref{building} ), and the identification of binaries (Sect.\,\ref{biniden}). Sect.\,\ref{tempe}  presents our determinations of effective temperature for single and binary targets. Finally, in Sect.\,\ref{conclu}  we summarize the results of this work and give indications on how to proceed in future binary searches on larger hot sds catalogues.

\section{Analysis}
\label{analysis}
\subsection{Object selection}
\label{sel}

The catalogue of hot sds compiled by \citet{Geier20} contains 5\,874 unique sources (2\,187 with spectroscopically-derived physical parameters taken from the literature), out of which 528 are reported as new hot sds. Also, 777 and 1\,365 sources were classified as binaries depending on whether a spectroscopic or a photometric approach was used, respectively. We use this catalogue as a reference throughout the present work.

\begin{figure}
        \includegraphics[width=\columnwidth]{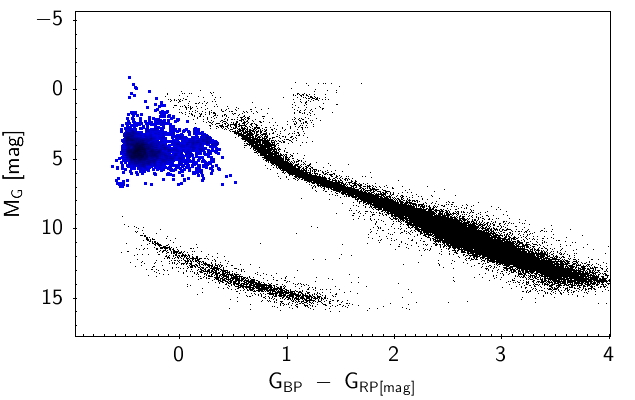}
   \caption{Colour-magnitude diagram built using \emph{Gaia} EDR3 sources with parallaxes larger than 10 mas, photometric errors in $G$, $G_\mathrm{BP}$ and $G_\mathrm{RP}$ less than 10 per cent, RUWE $<$ 1 to remove binaries 
   \citep{Arenou2018,Lindegren2018,Lindegren2021} and the colour excess factor applied following the approach described in \citet{Riello20} (black dots). The 3\,186 hot subdwarf stars selected following the criteria described in Sect. \ref{sel} are overplotted as blue dots.}


        \label{fig:hrd}
\end{figure}

We perform a first filtering of the catalogue keeping only those objects with low relative errors in magnitudes and parallaxes. We are interested in keeping sources with good values of distances as they will be used as filtering criterion in the determination of effective temperatures (Sect\,\ref{tempe}). With this purpose, we first cross-matched the 5\,874 objects of \citet{Geier20} with the \emph{Gaia} Early Data Release 3 (EDR3) catalogue \citep{Gaia2020} using a 5\,arcsec radius. We found 5\,862 sources with counterparts in \emph{Gaia} EDR3. Since the Geier's catalogue avoids the Galactic plane, the contamination from matching to other stars within this search radius can be considered as negligible. Anyway, in those cases where more than one counterpart exists in the search region, only the nearest one to the Geier's source was considered. Then, we kept only counterparts with relative errors of less than 10 per cent in $G$, $G_\mathrm{BP}$ and $G_\mathrm{RP}$ and less than 20 per cent in parallax. The absolute \emph{Gaia} magnitude in the $G$ band was estimated using

\begin{equation}
  M_G = G + 5 \log{\varpi} +5,   
\end{equation}

where $\varpi$ is the parallax in arcseconds. In our case, the inverse of the parallax is a reliable distance estimator because we kept only sources with relative errors in parallax lower than 20 per cent \citep{Luri18}. After this filtering, 3\,571 sources were kept. The great majority of the sources (all but four) were rejected because they did not fulfill the condition on parallaxes.
Then, we applied the color and absolute magnitude selection criteria described in \cite{Geier19} to define the expected locus for hot sds. In particular, we kept objects occupying the region -0.7 $<$ $G_\mathrm{BP}$-$G_\mathrm{RP} \leq$ 0.7 and -1.0 $<$ $M_\mathrm{G}$ $<$ 7.0 mag and fulfilling Eqs. (1)-(3) in that paper. This selection resulted in 3\,186 objects (Fig.\,\ref{fig:hrd}).

\subsection{SED building}
\label{building}

For the 3\,186 objects selected in the previous section, we used VOSA\footnote{http://svo2.cab.inta-csic.es/theory/vosa/} \citep{Bayo08} to build their SEDs. VOSA is a tool developed by the Spanish Virtual Observatory\footnote{http://svo.cab.inta-csic.es} designed to build the SEDs of thousands of objects at a time from a large number of photometric catalogues, ranging from the ultraviolet to the infrared. VOSA compares catalogue photometry with different collections of theoretical models and determines which model best reproduces the observed data following different statistical approaches. Physical parameters can then be estimated for each object from the model that best fits the data. 


In our analysis, the photometric information was obtained from the following catalogues: GALEX \citep{Bianchi00}, Pan-STARRS DR2 \citep{Magnier20}, \emph{Gaia} \citep{Gaia2020}, SDSS DR12 \citep{Alam15}, 2MASS-PSC  \citep{Skrutskie06}, and WISE \citep{Wright10}. 


\subsection{Binary identification}
\label{biniden}

VOSA also allows the identification of flux excess in a SED due to the presence of, for instance,
a disc, a nebula or a close companion. To this end, VOSA  first executes  an  iterative algorithm  which is  an extension  of  the method  described  in  \citet{Lada06}. Starting  at
$\lambda \geq 21500\,\textup{\r{A}}$, VOSA  computes the slope of the linear
regression  of   the  observational   SED  in  a   log\,$\nu  F_{\nu}$
vs.  log\,$\nu$  diagram.  This  slope is  recomputed  by  adding  new
infrared photometric points at every step.  If, in any of these steps,
the  slope  becomes  significantly  smaller than  the  one
expected from a  stellar photospheric emission, VOSA  flags the object
as potentially affected by excess and photometric points at longer
wavelengths are not taken into account  in the SED fitting process.

Once the SED fitting is  completed, VOSA performs a further refinement
of the excess estimation  by comparing, for each photometric point,
the observational flux  to the synthetic flux obtained  from the model
that best  fits the  data. Significant ($>$\,3$\sigma$)  deviations in
the observational flux are flagged by VOSA as potential IR excesses. A
detailed description  of how VOSA  manages the infrared excess  can be
found                   in                  the                   VOSA
documentation\footnote{https://bit.ly/2KRCv9x}.  

From the sample of 3\,186 objects, VOSA identifed 2\,469 single objects and 615 binary systems. The rest of the objects (102) were not classified because of their bad SED fitting due to, for instance, poor quality photometry or lack of enough photometric points. This gives a binarity ratio of 20 per cent (615 / (615+2\,469)), in agreement with \citet{Geier20} (13 per cent (777 / 5\,874) or 23 per cent (1\,365 / 5\,874), depending on if, in their study, spectroscopic of photometric binaries are considered). We note that our estimation can be biased by selection effects indicating just a lower limit to the real number of binaries as our detection of companions is restricted to not too cool objects (that would be out-shined by the hot sd in the optical regime), not too hot objects (that would out-shine the hot sd) as well as to systems where the difference in effective temperature between the hot sd and the companion is significant (systems with similar \teff\, will be seen as a single SED by VOSA). Also, objects without good photometric information in the 2MASS-WISE regime (which is where the flux excess indicating the presence of a companion occurs) could be wrongly classified as singles. In fact, 232 out of the 2\,469 objects classified as singles lie in this category. Assuming a binarity ratio of 20 per cent, this would imply that $\sim$ 50 sources classified as single could be real binaries. Moreover, hot sds showing large reflection effects will have associated SEDs with large photometric variability as the photometric data to construct the SED were obtained at different observing epochs. They will have a bad SED fitting lying, thus, among the 102 not classified objects. However, even in the unrealistic scenario in which the 102 objects with bad SED fitting are true reflection binaries, the binarity ratio will not change too much: from 20\% (615/(615+2\,469)) to 22.5\% (717/(717+2\,469)). Finally, wide binaries escape from our search as they do not present a composite SED, although \citet{Pelisoli20} suggested that only a small fraction of hot sds are in wide binaries.


\citet{Geier17} provide a classification of hot sds in three classes (sdOs, sdBs and sd+MS) following a purely empirical scheme based on photometry. \cite{Geier20} expands this methodology using SDSS \citep{Alam15}, APASS \citep{Henden16}, Pan-STARRS1 \citep{Chambers16} and SKyMapper \citep{Wolf18} data. The SDSS- and SkyMapper-based colour classes are regarded as the most trustworthy because the $u-g$ colour allows to distinguish between sdB and the hotter sdO types better than any other colour combination.

\begin{figure}
        \includegraphics[width=\columnwidth]{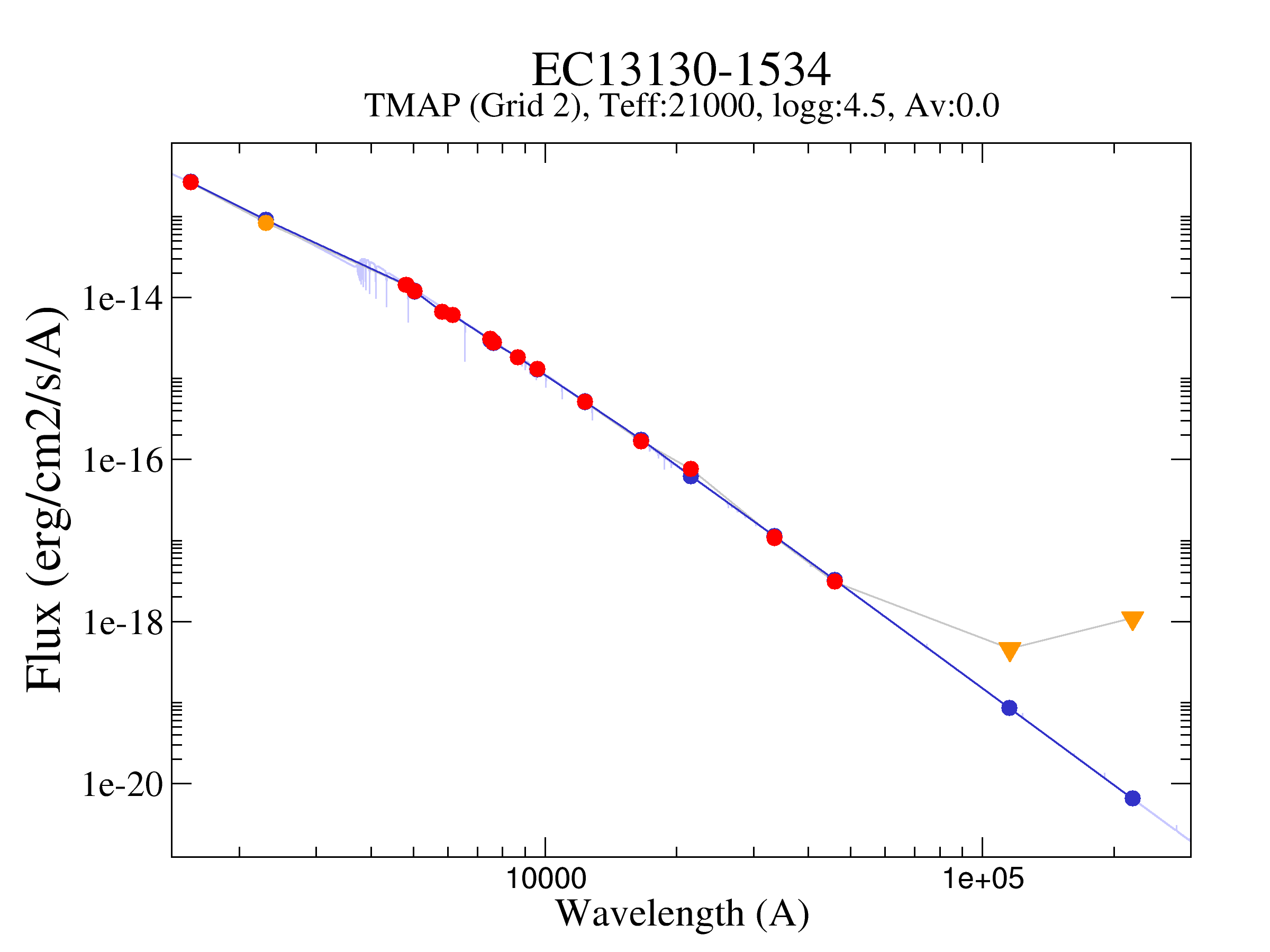}
        \includegraphics[width=\columnwidth]{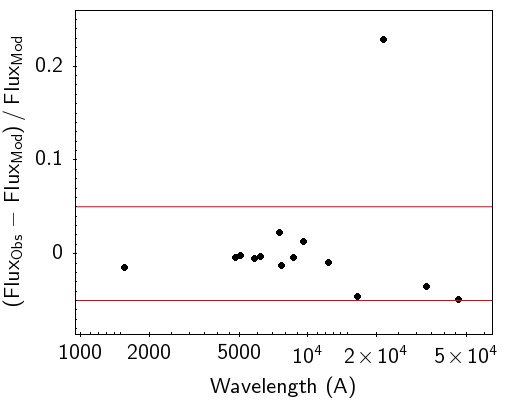}
   \caption{Top: Spectral energy distributions of EC13130-1534, photometrically classified as binary in \citet{Geier20} but whose model fitting with VOSA indicates that it is most likely a single star. The red dots represent the observed photometry while the blue spectrum indicates the theoretical model that fits best. The yellow inverted triangle indicates that the photometric value corresponds to an upper limit and, thus, is not taken into account in the fitting process. Bottom: Residuals of the fit for the good photometric points. Except for $K_{s}$, the differences between the observed and the model flux are below 5 per cent (red lines).}
 
        \label{fig:bin}
\end{figure}

Among our list of 3\,084 objects (2\,469 singles + 615 binaries), there are 548 photometrically classified as binaries by \citet {Geier20} using any of their colour-colour combinations. However we found that, for 202 of them (37 per cent), the SED can be better fitted (in terms of $\chi^{2}$) using a single object model (see Fig.\,\ref{fig:bin} as an example). Only 24 of them are spectroscopically confirmed as binaries in \citet{Geier20}, which reinforces the hypothesis of their single nature and indicates that VOSA offers a superior performance discriminating between singles and binaries than the photometric criteria used in \citet{Geier20}.

\begin{figure}
        \includegraphics[width=\columnwidth]{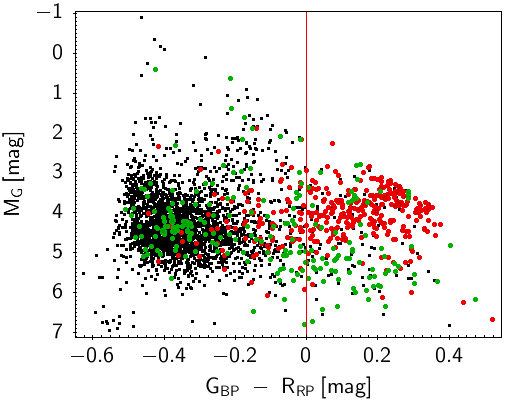}
    \caption{Colour-magnitude diagram showing the initial list of 3\,084 objects (black dots), the 548 objects classified as binaries by \citet{Geier20} (red dots) and the 202 objects classified as binaries by \citet{Geier20} but as single sources by VOSA (green dots). The vertical solid line represents the frontier between the binary  ($G_\mathrm{BP}$ - $G_\mathrm{RP}$ with positive values) and single ($G_\mathrm{BP}$ - $G_\mathrm{RP}$ with negative values) locii as described in \citet {Geier19}.}
\label{fig:bin0}
\end{figure}

As shown in Fig.\,\ref{fig:bin0}, the majority of these sources (141/202, 70 per cent) are located at $G_\mathrm{BP}$ - $G_\mathrm{RP}$ $ \leq$ 0.0, which is the locus for single hot sds according to \cite{Geier19}. If only SDSS photometry is used, then, 253 sources are classified as binaries, out of which 66 (26 per cent) are classified as single by VOSA. 54 of these 66 sources (89 per cent) have negative $G_\mathrm{BP}$ - $G_\mathrm{RP}$ values. 
Besides, among the 615 binary systems identified by VOSA, 269 objects were not reported as binaries in \citet{Geier20}.

\section{Effective temperature determination}
\label{tempe}
\subsection{Single SED objects}
\label{single}
For the 2\,469 objects classified as single by VOSA we attempted to estimate their effective temperatures. For this, we made use of the TMAP (Tubingen NLTE Model Atmosphere Package) collection of theoretical models \citep{Rauch03}. Hydrogen-only models in the 20\,000\,K $<$ \teff $<$ 150\,000\,K range were requested to TheoSSA\footnote{http://dc.zah.uni-heidelberg.de/theossa/q/web/form} and implemented in the theoretical spectra web server of Spanish Virtual Observatory portal\footnote{http://svo2.cab.inta-csic.es/theory/newov2} for their use by VOSA. In our analysis the range in surface gravity was restricted to log\,g: 4.5 -- 6\,dex, typical of hot subdwarfs.

Extinction can play an important role in shaping the SED in particular at short wavelengths where the peak of the SED is reached. If extinction is underestimated, the slope of the SED will appear flattened at short wavelengths and the derived effective temperature will be lower. To account for this effect and to minimize the extinction - effective temperature degeneracy in the SED fitting, we decided to leave extinction as a free parameter in the SED fitting process taking values in the range 0 $\leq$ Av $\leq$ 1\,mag and keep only objects at distances $<$ 1000 pc. VOSA estimates extinction using the laws described in \citet{Fitzpatrick99} (optical) and \citet{Indebetouw05} (infrared). Distances were obtained from \emph{Gaia} EDR3 in the way explained in Sect\,\ref{sel}.

\begin{figure}
        \includegraphics[width=\columnwidth]{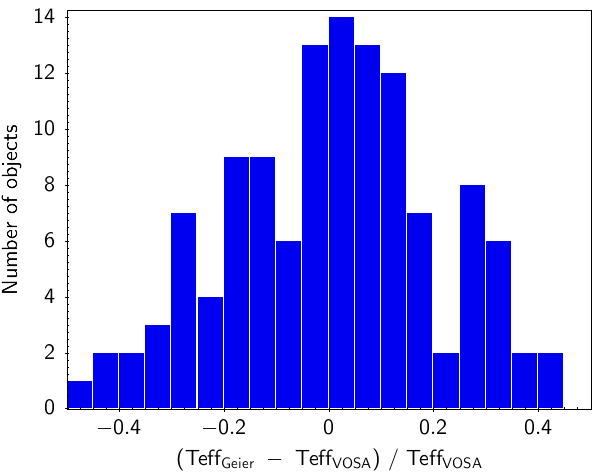}
    \caption{Relative differences between the spectroscopic temperatures given in \citet{Geier20} and those derived using VOSA for the 98 objects in common fulfilling the conditions described in Section \ref{single}. More than 70\% of the objects have relative errors below 20\% with VOSA temperatures being slightly lower than those given in \citet{Geier20}. This could be associated to an underestimation of the extinction factor, Av.}
   \label{fig:teff_comp}
\end{figure}

After this, we kept only the objects fulfilling the following filtering criteria: 

\begin{itemize}

\item Given the high temperatures of the hot sds, the accuracy of the values estimated from the SED fitting using VOSA is closely linked to the photometric coverage in the blue part of the electromagnetic spectrum. Therefore, we kept only objects having both GALEX FUV and NUV photometry. GALEX photometry was corrected using the expressions given in \citet[table 5]{Camarota14}\\
\item We selected objects with \teff $>$ 20\,000\,K, to avoid boundary issues as our grid of models starts at \teff\,= 20\,000\,K. \\
\item We selected objects with \teff $<$ 50\,000\,K. Higher temperatures will be affected by large uncertainties due to the lack of photometric information at wavelengths shorter than GALEX. According to the Wien's law, the SED of an object with \teff$ $ = 50\,000\,K peaks at 576 \AA, well beyond the GALEX FUV band ($\lambda_{mean}$ = 1546 \AA). Although this cut in effective temperature implies that hot sdOs will not be considered, this should not have a major impact in the final number of candidates since, as \citet{Heber16} pointed out, the large majority of hot sds have a temperature between 20\,000 and 40\,000\,K.\\

\item Good SED fitting (\emph{vgfb}$<$15). \emph{vgfb} is a modified $\chi^2$, internal to VOSA, and calculated by forcing $\Delta F_\mathrm{bol}$ to be, at least, 0.1 $\times$ F$_\mathrm{bol}$, where $\Delta F_\mathrm{bol}$ is the error in the total observed bolometric flux. This parameter is particularly useful when the photometric errors of any of the catalogues used to build the SED are underestimated. \emph{vgfb}$<$ 15 is a reliable indicator for a good fit. 
   \end{itemize}
   After applying the previous conditions, only 192 sources were kept, out of which, 98 have determinations of effective temperature in \citet{Geier20}. This subset of 98 objects was used to assess the reliability and robustness of the effective temperatures calculated with VOSA.
   Fig.\,\ref{fig:teff_comp} shows the relative differences between these values and those derived with VOSA. The agreement is good with VOSA temperatures being slightly lower than those in \citet{Geier20}. Information on the 192 objects for which effective temperatures were estimated can be found at the SVO archive of hot sds (see Appendix).
   

\subsection{Binary SED objects}
\label{binary}

If there is a difference in \teff$ $ between the hot sd and the main sequence companion (with the hot sd mostly contributing at shorter wavelengths while the companion dominates at longer wavelengths), VOSA is able to decompose the observed SED. Physical parameters can then be estimated for each one of the components by taking advantage of the two-body fitting algorithm implemented in VOSA (see the VOSA documentation\footnote{http://svo2.cab.inta-csic.es/theory/vosa50/helpw4.php?otype=starnew\\
\&what=action=help\&what=fitbin} for a detailed description on how the two-body SED fitting is made). 

To ensure the reliability of the results provided by our methodology, we first applied it to known composite hot sds with spectroscopically derived effective temperatures. In particular we used the list of objects compiled in \citet{Girven12} and \cite{Nemeth12}, which provide effective temperatures for both components for 227 and 29 objects, respectively. After applying the restrictions in terms of distances (parallax $>1$ and  relative error $<$ 20 per cent), GALEX photometry and good SED fitting, we ended up with 6 and 11 objects, respectively. 
SEDs were fitted using the TMAP grid of models for the hot sd component and BT-Settl models \citep{allard12} for the cooler companion. The ranges of effective temperatures and surface gravities  were set to 20\,000 -- 50\,000\,K and 4.5 -- 6\,dex and 2\,000\,K -- 19\,000\,K and 4 -- 6\,dex, for the hot sd and the companion, respectively. 

In Fig.\,\ref{fig:doubles_lit}\,we compare the effective temperatures provided by VOSA with the spectroscopic determinations for those 17 systems. We find that there are no systematic differences. No companions cooler than K-types are detected because they are completely out-shined by the hot sd. Similarly, companions hotter than A-types would dominate the SED flux, making the hot sd unseen. The spread in temperatures, in particular for the hot sds, can be ascribed to the following reasons:

\begin{figure}
        \includegraphics[width=\columnwidth]{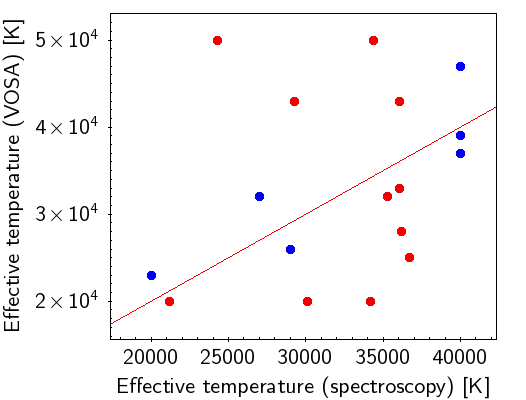}
        \includegraphics[width=\columnwidth]{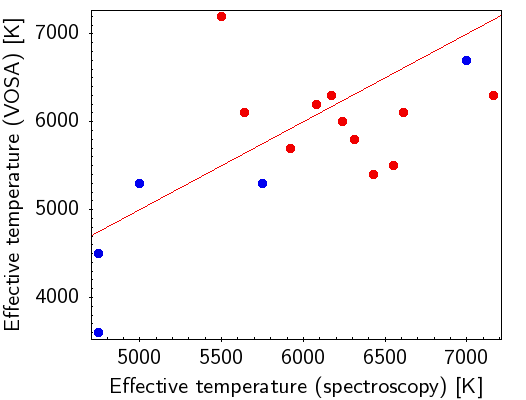}
    \caption{Comparison between the effective temperatures estimated with VOSA and the corresponding values taken from \citet{Girven12} (blue) and \citet{Nemeth12} (red) for the hot sds (top) and the cool companions (bottom).}

        \label{fig:doubles_lit}
\end{figure}

\begin{figure}
        \includegraphics[width=\columnwidth]{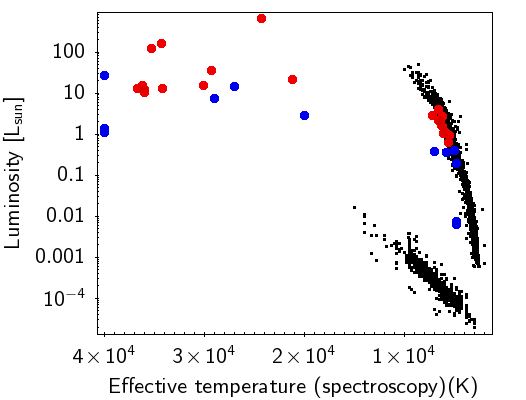}
    \caption{Position in a $L$ -- $T_\mathrm{eff}$ diagram of the \citet{Girven12} (blue) and \citet{Nemeth12} (red) objects analysed in Sec\,\ref{binary}. No subgiant/giant companions have been identified in our sample. Black dots are a subsample of the small black dots shown in Fig.\,\ref{fig:hrd} representing the main sequence and the white dwarf cooling track.
     \label{fig:lum}}
   \end{figure}

\begin{itemize}

\item As previously mentioned, extinction is left as a free parameter in the VOSA two-body fitting process. $A_{V}$ ranges from 0.0 to 1.0\,mag in steps of 0.05\,mag, according to the selection in distance ($<$ 1Kpc). A wrong estimation of the extinction would lead to a wrong estimation of the effective temperature, in the sense that systematically low values of $A_{V}$ would give systematically low values of temperatures. \\

    
    \item We assume a fixed metallicity (solar) both for the hot sd and the companion. If one (or both) component was a metal-poor object, the line-blanketing in the ultraviolet would be less pronounced which would translate into an increase of the flux at short wavelengths. The composite SED would show a flux excess in the GALEX range leading to a bad fitting to solar metallicity models in this regime. However, these objects would fall in the category \emph{"binaries with bad SED fitting"} and, thus, their temperatures would have not been estimated. Moreover, as we are selecting objects at less that 1 Kpc, the fraction of thick disc and halo stars is negligible. Therefore, we can discard metallicity as one of the main effects causing the spread in effective temperatures.\\

    
     \item \citet{Girven12} pointed out that the spectroscopic analysis of the composite hot sd is often carried out using a single-star model. However, the companion may have a significant contribution to the absorption lines used in the analysis resulting, thus, in a biased estimation of the system parameters, in particular of the spectroscopically derived effective temperatures.\\
     
    \item In all the binary systems studied in this work, the companions have been assumed to be main sequence stars. However, examples of giants and subgiants companions to hot sds have been previously reported \citep{Heber02}. To discard this possibility we have placed the hot sds and their companions in a luminosity -- effective temperature diagram. Fig.\,\ref{fig:lum} shows the position of the hot sds and their companions included in Fig.\,\ref{fig:doubles_lit}\,. We see how all our companions lie on or just below the main sequence, which discards the presence of subgiant/giant companions. We note that the range validity of the computed luminosity values is severely hampered by the small sample size (17 systems) available. Besides, the lack of photometric data bluer than GALEX data largely influences real uncertainties associated to temperature and, in turn, luminosity determinations.
Other considerations, such as ageing or He content could set hot sds at higher (or lower) luminosity values \citep{Kawka15}
\\
     
    \item As \citet{Rauch08} suggested, different physical approaches in the model atmosphere codes, in particular in the UV range,  can translate into non negligible differences in the values of the derived physical parameters. \citet{Girven12} uses \citet{Heber00} while \citet{Nemeth12} relies on \citet{Hubeny95} models. A detailed comparison between the outputs of these theoretical models -- as well as the different methodologies used in \citet{Girven12} and  \citet{Nemeth12} -- goes beyond the scope of this paper.  

\end{itemize}

\begin{figure}
      \includegraphics[width=\columnwidth]{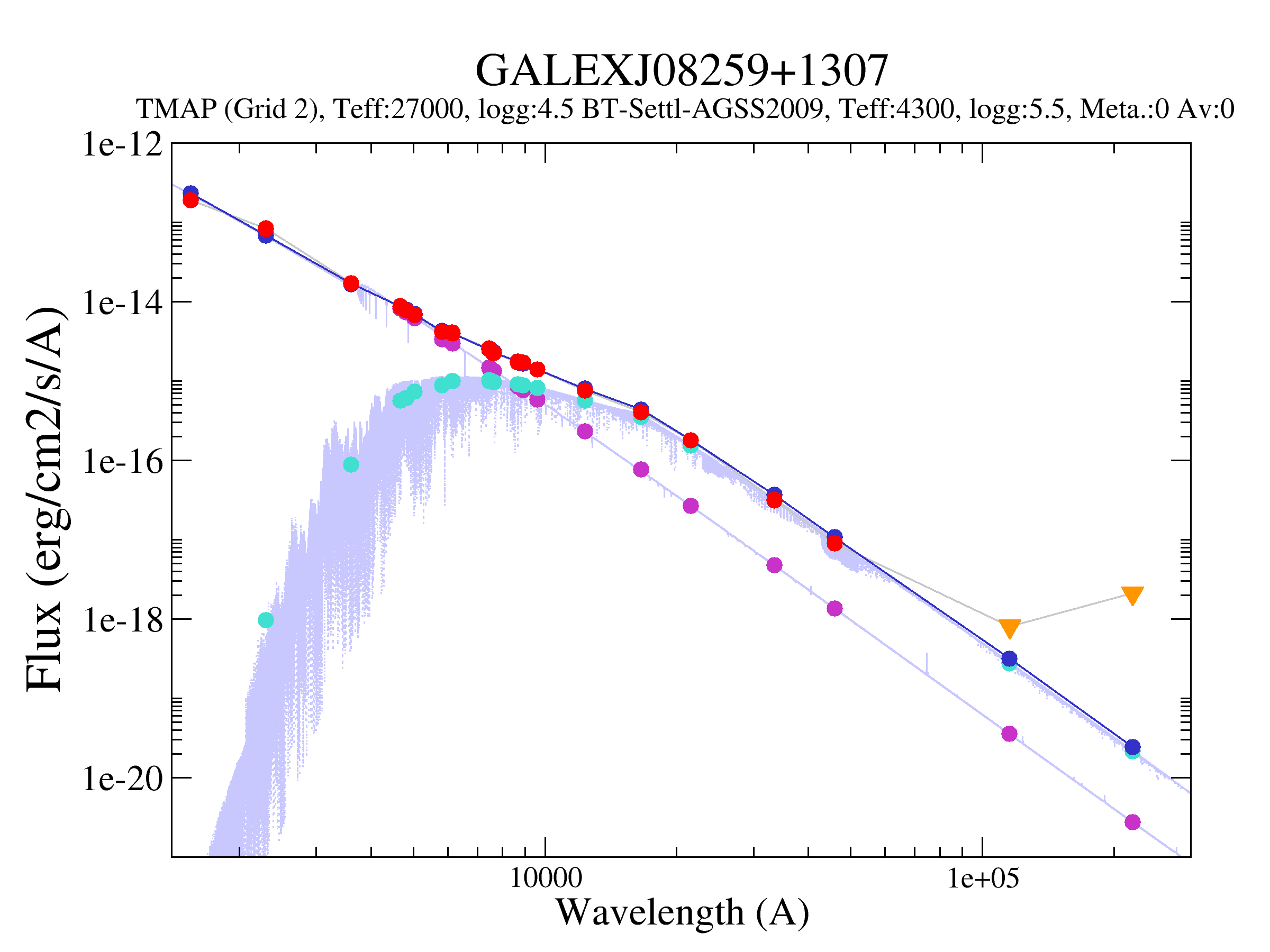}
       \includegraphics[width=\columnwidth]{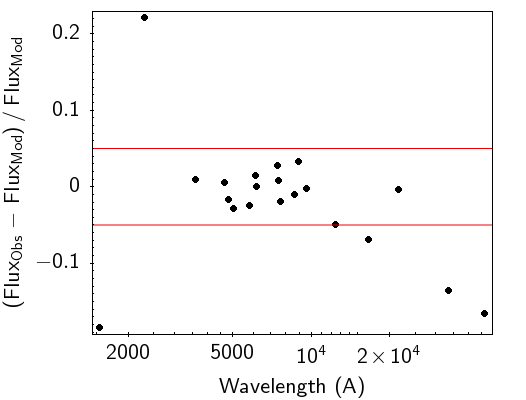}
    \caption{Top: Example of a two-body SED fitting using VOSA. The purple and green dots represent the TMAP and BT-Settl photometry of the model that fits best, respectively. The corresponding theoretical spectra are both plotted in purple. Red dots are the observed photometric points while the blue line and dots indicate the composite model that best fits the data. The yellow inverted triangles indicate that the photometric values correspond to upper limits and, thus, are not taken into account in the fitting process. Bottom: Residuals of the fit for the good photometric
points (black dots). Red lines indicate a difference of 5 per cent between the observed and the model flux.}

        \label{fig:SED_double}
\end{figure}

\begin{figure}
      \includegraphics[width=\columnwidth]{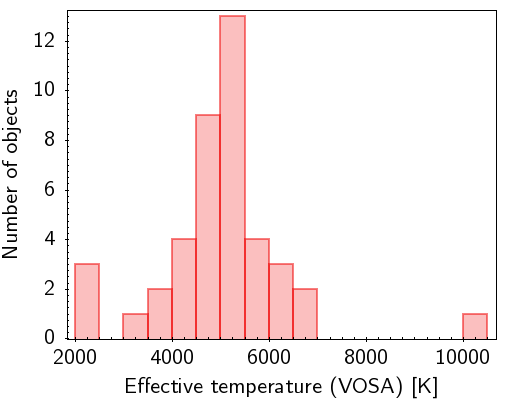}
    \caption{Effective temperatures of the cool companions of the 42 hot sds binaries for which physical parameters were derived from the SED two-body fitting. See Sect.\,\ref{binary} for details.}

        \label{fig:hist_comp}
\end{figure}

Once we have confirmed that our methodology provide reliable values of \teff, we applied it to the 615 objects included in \citet{Geier20} and classified as binaries by VOSA. After applying the filtering criteria described in Sect.\,\ref{single} (i.e., conditions in range of temperatures, distances and UV photometry), only 70 sources were kept. These objects were visually inspected and only 42 of them showed a good SED two-body fitting (Fig.\,\ref{fig:SED_double}, as an example). The effective temperature distribution of the companions is shown in Fig.\,\ref{fig:hist_comp}. We can see how the peak in the distribution is reached at $T_\mathrm{eff}$ = 5\,000 - 5\,500\,K (G7V\,-\,K2V), while 75 per cent of the objects lie in the range 4\,000\,-\,6\,000\,K (F9V\,-K8V). Particularly interesting are the three objects with substellar temperatures ($T_\mathrm{eff}$ $<$ 2\,500\,K) which would require further and more detailed analysis to confirm their substellar nature. Information on these 42 objects can be found in the SVO archive of hot sds (see Appendix). 


%

\section{Conclusions}
\label{conclu}

A substantial fraction of hot subdwarfs reside in binaries -- see \citet {Kawka15, Pelisoli20}, and references therein. These are important objects to constrain the different evolution scenarios proposed to explain the loss of the hydrogen envelope during the red giant branch phase. In this work we have described a method to identify hot sd binaries with  the  help  of  the  VOSA  Virtual  Observatory  tool. The method consists on three steps: i) Building of the spectral energy distribution from the ultraviolet to the infrared. ii) Identification of binaries based on flux excess towards redder bands. iii) Physical parameter estimation for both single and composite hot subdwarfs from the model that best fits. 

Our procedure was checked against the hot subdwarf catalogue of \citet{Geier20}, the latest and most complete catalogue of hot sds presently available. From a sample of the 3\,186 objects with good photometric and astrometric information in \emph{Gaia} EDR3 and lying on the expected locus for hot sds, VOSA identifed 2\,469 single objects and 615 binary systems, showing a superior performance to the colour-based criterion proposed in \citet{Geier20} and finding 269 binaries not reported in that paper. The rest of objects (102) were not classified by VOSA due to their bad SED fitting. This gives a binary ratio of 20 per cent, a number that should be regarded as a lower limit for the binarity incidence which, in any case, lies within expectations. For the 2\,469 single hot sds, we computed temperatures, luminosities and radii for 192 objects. Likewise, out of the 615 objects identified as binaries by VOSA, physical parameters could be reliably estimated for 42 of them using a two-body SED fitting.   

According to
\citet{Nemeth20}, binary companions to hot sds could be distributed
as follows: 13-17 per cent white dwarf, 26-30 per cent low mass MS or near
the substellar limit, and about 20 per cent more massive MS. Besides,
MS+hot sd segregate into two main groups: i) systems formed
through common-envelope evolution would display orbital periods
of less than 30 days and a low mass companion; ii) systems formed
through Roche-lobe overflow have periods in the range of 400-1500
days and more massive F-G-K companions. Ideally, a careful
spectral decomposition analysis would also reveal reflection effects for
the shortest-period systems, chemical compositions, masses, ages
or rotational properties. In this context, our results provide an excellent bona-fide hot subdwarf binary candidates sample to be spectroscopically followed up by potentially interested researchers in the field.

Also, the application of Artificial Intelligence (AI) techniques to larger hot subdwarf candidate datasets, such as for example the all-sky catalogue of 39\,800 hot subluminous star candidates \citep{Geier19}, and the use of new \emph{Gaia} data releases could be considered for future studies. Deep Learning techniques such as autoencoders and convolutional networks have proven useful in classifying objects with incomplete or noisy data, and can provide practical classifications if sufficient reference data are available to train the networks.

\section*{Acknowledgements}

This work has been funded by
RTI2018-095076-B-C22 and also by
MCIN/AEI/10.13039/501100011033 through grant PID2020-112949GB-I00 and MDM-2017-0737 at Centro de Astrobiolog\'{i}a (CSIC-INTA), Unidad de Excelencia Mar\'{i}a de Maeztu. We acknowledge support from CIGUS CITIC, funded by Xunta de Galicia and the European Union (FEDER Galicia 2014-2020 Program) through grant ED431G 2019/01; research consolidation grant ED431B 2021/36; and scholarship from Xunta de Galicia 620 and the European Union (European Social Fund - ESF) ED481A-2019/155. A.A. acknowledges support from Government of Comunidad Autónoma de Madrid (Spain) through postdoctoral grant ‘Atracción de Talento Investigador’ 2018-T2/TIC-11697. This publication makes use of VOSA, developed under the Spanish Virtual Observatory project. This research has made use of Aladin sky atlas developed at CDS, Strasbourg Observatory, France \citep{Bonnarel00, Boch14}. Vizier \citep{Ochsenbein00}, TOPCAT \citep{Taylor05} and STILTS \citep{Taylor06} have also been widely used in this paper.

\section{Data availability}
The data underlying this article are available in the SVO archive of hot subdwarfs at http://svocats.cab.inta-csic.es/hsa2/



\bibliographystyle{mnras}
\bibliography{mnras} 

\begin{thebibliography}{}
\makeatletter
\relax
\def\mn@urlcharsother{\let\do\@makeother \do\$\do\&\do\#\do\^\do\_\do\%\do\~}
\def\mn@doi{\begingroup\mn@urlcharsother \@ifnextchar [ {\mn@doi@}
  {\mn@doi@[]}}
\def\mn@doi@[#1]#2{\def\@tempa{#1}\ifx\@tempa\@empty \href
  {http://dx.doi.org/#2} {doi:#2}\else \href {http://dx.doi.org/#2} {#1}\fi
  \endgroup}
\def\mn@eprint#1#2{\mn@eprint@#1:#2::\@nil}
\def\mn@eprint@arXiv#1{\href {http://arxiv.org/abs/#1} {{\tt arXiv:#1}}}
\def\mn@eprint@dblp#1{\href {http://dblp.uni-trier.de/rec/bibtex/#1.xml}
  {dblp:#1}}
\def\mn@eprint@#1:#2:#3:#4\@nil{\def\@tempa {#1}\def\@tempb {#2}\def\@tempc
  {#3}\ifx \@tempc \@empty \let \@tempc \@tempb \let \@tempb \@tempa \fi \ifx
  \@tempb \@empty \def\@tempb {arXiv}\fi \@ifundefined
  {mn@eprint@\@tempb}{\@tempb:\@tempc}{\expandafter \expandafter \csname
  mn@eprint@\@tempb\endcsname \expandafter{\@tempc}}}

\bibitem[\protect\citeauthoryear{{Alam} et~al.,}{{Alam} et~al.}{2015}]{Alam15}
{Alam} S.,  et~al., 2015, \mn@doi [\apjs] {10.1088/0067-0049/219/1/12}, \href
  {https://ui.adsabs.harvard.edu/abs/2015ApJS..219...12A} {219, 12}

\bibitem[\protect\citeauthoryear{{Allard}, {Homeier}  \& {Freytag}}{{Allard}
  et~al.}{2012}]{allard12}
{Allard} F.,  {Homeier} D.,   {Freytag} B.,  2012, \mn@doi [Philosophical
  Transactions of the Royal Society of London Series A]
  {10.1098/rsta.2011.0269}, \href
  {https://ui.adsabs.harvard.edu/abs/2012RSPTA.370.2765A} {370, 2765}

\bibitem[\protect\citeauthoryear{{Arenou} et~al.,}{{Arenou}
  et~al.}{2018}]{Arenou2018}
{Arenou} F.,  et~al., 2018, \mn@doi [\aap] {10.1051/0004-6361/201833234}, \href
  {https://ui.adsabs.harvard.edu/abs/2018A&A...616A..17A} {616, A17}

\bibitem[\protect\citeauthoryear{{Bayo}, {Rodrigo}, {Barrado Y Navascu{\'e}s},
  {Solano}, {Guti{\'e}rrez}, {Morales-Calder{\'o}n}  \& {Allard}}{{Bayo}
  et~al.}{2008}]{Bayo08}
{Bayo} A.,  {Rodrigo} C.,  {Barrado Y Navascu{\'e}s} D.,  {Solano} E.,
  {Guti{\'e}rrez} R.,  {Morales-Calder{\'o}n} M.,   {Allard} F.,  2008, \mn@doi
  [\aap] {10.1051/0004-6361:200810395}, \href
  {https://ui.adsabs.harvard.edu/#abs/2008A&A...492..277B} {492, 277}

\bibitem[\protect\citeauthoryear{{Bianchi} \& {GALEX Team}}{{Bianchi} \& {GALEX
  Team}}{2000}]{Bianchi00}
{Bianchi} L.,  {GALEX Team} 2000, \memsai, \href
  {https://ui.adsabs.harvard.edu/abs/2000MmSAI..71.1123B} {71, 1123}

\bibitem[\protect\citeauthoryear{{Boch} \& {Fernique}}{{Boch} \&
  {Fernique}}{2014}]{Boch14}
{Boch} T.,  {Fernique} P.,  2014, in {Manset} N.,  {Forshay} P.,  eds,
  Astronomical Society of the Pacific Conference Series Vol. 485, Astronomical
  Data Analysis Software and Systems XXIII. p.~277

\bibitem[\protect\citeauthoryear{{Bonnarel} et~al.,}{{Bonnarel}
  et~al.}{2000}]{Bonnarel00}
{Bonnarel} F.,  et~al., 2000, \mn@doi [\aaps] {10.1051/aas:2000331}, \href
  {https://ui.adsabs.harvard.edu/abs/2000A&AS..143...33B} {143, 33}

\bibitem[\protect\citeauthoryear{{Camarota} \& {Holberg}}{{Camarota} \&
  {Holberg}}{2014}]{Camarota14}
{Camarota} L.,  {Holberg} J.~B.,  2014, \mn@doi [\mnras]
  {10.1093/mnras/stt2422}, \href
  {https://ui.adsabs.harvard.edu/abs/2014MNRAS.438.3111C} {438, 3111}

\bibitem[\protect\citeauthoryear{{Chambers} et~al.,}{{Chambers}
  et~al.}{2016}]{Chambers16}
{Chambers} K.~C.,  et~al., 2016, arXiv e-prints, \href
  {https://ui.adsabs.harvard.edu/abs/2016arXiv161205560C} {p. arXiv:1612.05560}

\bibitem[\protect\citeauthoryear{{D'Cruz}, {Dorman}, {Rood}  \&
  {O'Connell}}{{D'Cruz} et~al.}{1996}]{Cruz96}
{D'Cruz} N.~L.,  {Dorman} B.,  {Rood} R.~T.,   {O'Connell} R.~W.,  1996,
  \mn@doi [\apj] {10.1086/177515}, \href
  {https://ui.adsabs.harvard.edu/abs/1996ApJ...466..359D} {466, 359}

\bibitem[\protect\citeauthoryear{{Drilling}, {Jeffery}, {Heber}, {Moehler}  \&
  {Napiwotzki}}{{Drilling} et~al.}{2013}]{Drilling13}
{Drilling} J.~S.,  {Jeffery} C.~S.,  {Heber} U.,  {Moehler} S.,   {Napiwotzki}
  R.,  2013, \mn@doi [\aap] {10.1051/0004-6361/201219433}, \href
  {https://ui.adsabs.harvard.edu/abs/2013A&A...551A..31D} {551, A31}

\bibitem[\protect\citeauthoryear{{Fitzpatrick}}{{Fitzpatrick}}{1999}]{Fitzpatrick99}
{Fitzpatrick} E.~L.,  1999, \mn@doi [\pasp] {10.1086/316293}, \href
  {http://adsabs.harvard.edu/abs/1999PASP..111...63F} {111, 63}

\bibitem[\protect\citeauthoryear{{Gaia Collaboration}}{{Gaia
  Collaboration}}{2020}]{Gaia2020}
{Gaia Collaboration} 2020, VizieR Online Data Catalog, \href
  {https://ui.adsabs.harvard.edu/abs/2020yCat.1350....0G} {p. I/350}

\bibitem[\protect\citeauthoryear{{Geier}}{{Geier}}{2020}]{Geier20}
{Geier} S.,  2020, \mn@doi [\aap] {10.1051/0004-6361/202037526}, \href
  {https://ui.adsabs.harvard.edu/abs/2020A&A...635A.193G} {635, A193}

\bibitem[\protect\citeauthoryear{{Geier} et~al.,}{{Geier}
  et~al.}{2011}]{Geier11}
{Geier} S.,  et~al., 2011, \mn@doi [\aap] {10.1051/0004-6361/201015316}, \href
  {https://ui.adsabs.harvard.edu/abs/2011A&A...530A..28G} {530, A28}

\bibitem[\protect\citeauthoryear{{Geier} et~al.,}{{Geier}
  et~al.}{2015}]{Geier15}
{Geier} S.,  et~al., 2015, \mn@doi [\aap] {10.1051/0004-6361/201525666}, \href
  {https://ui.adsabs.harvard.edu/abs/2015A&A...577A..26G} {577, A26}

\bibitem[\protect\citeauthoryear{{Geier}, {{\O}stensen}, {Nemeth}, {Gentile
  Fusillo}, {G{\"a}nsicke}, {Telting}, {Green}  \& {Schaffenroth}}{{Geier}
  et~al.}{2017}]{Geier17}
{Geier} S.,  {{\O}stensen} R.~H.,  {Nemeth} P.,  {Gentile Fusillo} N.~P.,
  {G{\"a}nsicke} B.~T.,  {Telting} J.~H.,  {Green} E.~M.,   {Schaffenroth} J.,
  2017, \mn@doi [\aap] {10.1051/0004-6361/201630135}, \href
  {https://ui.adsabs.harvard.edu/abs/2017A&A...600A..50G} {600, A50}

\bibitem[\protect\citeauthoryear{{Geier}, {Raddi}, {Gentile Fusillo}  \&
  {Marsh}}{{Geier} et~al.}{2019}]{Geier19}
{Geier} S.,  {Raddi} R.,  {Gentile Fusillo} N.~P.,   {Marsh} T.~R.,  2019,
  \mn@doi [\aap] {10.1051/0004-6361/201834236}, \href
  {https://ui.adsabs.harvard.edu/abs/2019A&A...621A..38G} {621, A38}

\bibitem[\protect\citeauthoryear{{Girven} et~al.,}{{Girven}
  et~al.}{2012}]{Girven12}
{Girven} J.,  et~al., 2012, \mn@doi [\mnras]
  {10.1111/j.1365-2966.2012.21415.x}, \href
  {https://ui.adsabs.harvard.edu/abs/2012MNRAS.425.1013G} {425, 1013}

\bibitem[\protect\citeauthoryear{{Heber}}{{Heber}}{2009}]{Heber09}
{Heber} U.,  2009, \mn@doi [\araa] {10.1146/annurev-astro-082708-101836}, \href
  {https://ui.adsabs.harvard.edu/abs/2009ARA&A..47..211H} {47, 211}

\bibitem[\protect\citeauthoryear{{Heber}}{{Heber}}{2016}]{Heber16}
{Heber} U.,  2016, \mn@doi [\pasp] {10.1088/1538-3873/128/966/082001}, \href
  {https://ui.adsabs.harvard.edu/abs/2016PASP..128h2001H} {128, 082001}

\bibitem[\protect\citeauthoryear{{Heber}, {Reid}  \& {Werner}}{{Heber}
  et~al.}{2000}]{Heber00}
{Heber} U.,  {Reid} I.~N.,   {Werner} K.,  2000, \aap, \href
  {https://ui.adsabs.harvard.edu/abs/2000A&A...363..198H} {363, 198}

\bibitem[\protect\citeauthoryear{{Heber}, {Moehler}, {Napiwotzki}, {Thejll}  \&
  {Green}}{{Heber} et~al.}{2002}]{Heber02}
{Heber} U.,  {Moehler} S.,  {Napiwotzki} R.,  {Thejll} P.,   {Green} E.~M.,
  2002, \mn@doi [\aap] {10.1051/0004-6361:20020127}, \href
  {https://ui.adsabs.harvard.edu/abs/2002A&A...383..938H} {383, 938}

\bibitem[\protect\citeauthoryear{{Henden}, {Templeton}, {Terrell}, {Smith},
  {Levine}  \& {Welch}}{{Henden} et~al.}{2016}]{Henden16}
{Henden} A.~A.,  {Templeton} M.,  {Terrell} D.,  {Smith} T.~C.,  {Levine} S.,
  {Welch} D.,  2016, VizieR Online Data Catalog, \href
  {https://ui.adsabs.harvard.edu/abs/2016yCat.2336....0H} {p. II/336}

\bibitem[\protect\citeauthoryear{{Hubeny} \& {Lanz}}{{Hubeny} \&
  {Lanz}}{1995}]{Hubeny95}
{Hubeny} I.,  {Lanz} T.,  1995, \mn@doi [\apj] {10.1086/175226}, \href
  {https://ui.adsabs.harvard.edu/abs/1995ApJ...439..875H} {439, 875}

\bibitem[\protect\citeauthoryear{{Indebetouw} et~al.,}{{Indebetouw}
  et~al.}{2005}]{Indebetouw05}
{Indebetouw} R.,  et~al., 2005, \mn@doi [\apj] {10.1086/426679}, \href
  {http://adsabs.harvard.edu/abs/2005ApJ...619..931I} {619, 931}

\bibitem[\protect\citeauthoryear{{Kawka}, {Vennes}, {O'Toole}, {N{\'e}meth},
  {Burton}, {Kotze}  \& {Buckley}}{{Kawka} et~al.}{2015}]{Kawka15}
{Kawka} A.,  {Vennes} S.,  {O'Toole} S.,  {N{\'e}meth} P.,  {Burton} D.,
  {Kotze} E.,   {Buckley} D.~A.~H.,  2015, \mn@doi [\mnras]
  {10.1093/mnras/stv821}, \href
  {https://ui.adsabs.harvard.edu/abs/2015MNRAS.450.3514K} {450, 3514}

\bibitem[\protect\citeauthoryear{{Kepler} et~al.,}{{Kepler}
  et~al.}{2015}]{Kepler15}
{Kepler} S.~O.,  et~al., 2015, \mn@doi [\mnras] {10.1093/mnras/stu2388}, \href
  {https://ui.adsabs.harvard.edu/abs/2015MNRAS.446.4078K} {446, 4078}

\bibitem[\protect\citeauthoryear{{Kepler} et~al.,}{{Kepler}
  et~al.}{2016}]{Kepler16}
{Kepler} S.~O.,  et~al., 2016, \mn@doi [\mnras] {10.1093/mnras/stv2526}, \href
  {https://ui.adsabs.harvard.edu/abs/2016MNRAS.455.3413K} {455, 3413}

\bibitem[\protect\citeauthoryear{{Kilkenny}, {Heber}  \& {Drilling}}{{Kilkenny}
  et~al.}{1988}]{Kilkenny88}
{Kilkenny} D.,  {Heber} U.,   {Drilling} J.~S.,  1988, South African
  Astronomical Observatory Circular, \href
  {https://ui.adsabs.harvard.edu/abs/1988SAAOC..12....1K} {12, 1}

\bibitem[\protect\citeauthoryear{{Lada} et~al.,}{{Lada} et~al.}{2006}]{Lada06}
{Lada} C.~J.,  et~al., 2006, \mn@doi [\aj] {10.1086/499808}, \href
  {http://adsabs.harvard.edu/abs/2006AJ....131.1574L} {131, 1574}

\bibitem[\protect\citeauthoryear{{Lei}, {Zhao}, {N{\'e}meth}  \& {Zhao}}{{Lei}
  et~al.}{2019}]{Lei19}
{Lei} Z.,  {Zhao} J.,  {N{\'e}meth} P.,   {Zhao} G.,  2019, \mn@doi [\apj]
  {10.3847/1538-4357/ab2edc}, \href
  {https://ui.adsabs.harvard.edu/abs/2019ApJ...881..135L} {881, 135}

\bibitem[\protect\citeauthoryear{{Lindegren} et~al.,}{{Lindegren}
  et~al.}{2018}]{Lindegren2018}
{Lindegren} L.,  et~al., 2018, \mn@doi [\aap] {10.1051/0004-6361/201832727},
  \href {https://ui.adsabs.harvard.edu/abs/2018A&A...616A...2L} {616, A2}

\bibitem[\protect\citeauthoryear{{Lindegren} et~al.,}{{Lindegren}
  et~al.}{2021}]{Lindegren2021}
{Lindegren} L.,  et~al., 2021, \mn@doi [\aap] {10.1051/0004-6361/202039709},
  \href {https://ui.adsabs.harvard.edu/abs/2021A&A...649A...2L} {649, A2}

\bibitem[\protect\citeauthoryear{{Luri} et~al.,}{{Luri} et~al.}{2018}]{Luri18}
{Luri} X.,  et~al., 2018, \mn@doi [\aap] {10.1051/0004-6361/201832964}, \href
  {https://ui.adsabs.harvard.edu/#abs/2018A&A...616A...9L} {616, A9}

\bibitem[\protect\citeauthoryear{{Magnier} et~al.,}{{Magnier}
  et~al.}{2020}]{Magnier20}
{Magnier} E.~A.,  et~al., 2020, \mn@doi [\apjs] {10.3847/1538-4365/abb82a},
  \href {https://ui.adsabs.harvard.edu/abs/2020ApJS..251....6M} {251, 6}

\bibitem[\protect\citeauthoryear{{N{\'e}meth}}{{N{\'e}meth}}{2020}]{Nemeth20}
{N{\'e}meth} P.,  2020, \mn@doi [Contributions of the Astronomical Observatory
  Skalnate Pleso] {10.31577/caosp.2020.50.2.546}, \href
  {https://ui.adsabs.harvard.edu/abs/2020CoSka..50..546N} {50, 546}

\bibitem[\protect\citeauthoryear{{N{\'e}meth}, {Kawka}  \&
  {Vennes}}{{N{\'e}meth} et~al.}{2012}]{Nemeth12}
{N{\'e}meth} P.,  {Kawka} A.,   {Vennes} S.,  2012, \mn@doi [\mnras]
  {10.1111/j.1365-2966.2012.22009.x}, \href
  {https://ui.adsabs.harvard.edu/abs/2012MNRAS.427.2180N} {427, 2180}

\bibitem[\protect\citeauthoryear{{Ochsenbein}, {Bauer}  \&
  {Marcout}}{{Ochsenbein} et~al.}{2000}]{Ochsenbein00}
{Ochsenbein} F.,  {Bauer} P.,   {Marcout} J.,  2000, \mn@doi [\aaps]
  {10.1051/aas:2000169}, \href
  {http://adsabs.harvard.edu/abs/2000A%26AS..143...23O} {143, 23}

\bibitem[\protect\citeauthoryear{{Oreiro}, {Rodr{\'\i}guez-L{\'o}pez},
  {Solano}, {Ulla}, {{\O}stensen}  \& {Garc{\'\i}a-Torres}}{{Oreiro}
  et~al.}{2011}]{Oreiro2011}
{Oreiro} R.,  {Rodr{\'\i}guez-L{\'o}pez} C.,  {Solano} E.,  {Ulla} A.,
  {{\O}stensen} R.,   {Garc{\'\i}a-Torres} M.,  2011, \mn@doi [\aap]
  {10.1051/0004-6361/201016324}, \href
  {https://ui.adsabs.harvard.edu/abs/2011A&A...530A...2O} {530, A2}

\bibitem[\protect\citeauthoryear{{Pelisoli}, {Vos}, {Geier}, {Schaffenroth}  \&
  {Baran}}{{Pelisoli} et~al.}{2020}]{Pelisoli20}
{Pelisoli} I.,  {Vos} J.,  {Geier} S.,  {Schaffenroth} V.,   {Baran} A.~S.,
  2020, \mn@doi [\aap] {10.1051/0004-6361/202038473}, \href
  {https://ui.adsabs.harvard.edu/abs/2020A&A...642A.180P} {642, A180}

\bibitem[\protect\citeauthoryear{{P{\'e}rez-Fern{\'a}ndez}, {Ulla}, {Solano},
  {Oreiro}  \& {Rodrigo}}{{P{\'e}rez-Fern{\'a}ndez} et~al.}{2016}]{Perez16}
{P{\'e}rez-Fern{\'a}ndez} E.,  {Ulla} A.,  {Solano} E.,  {Oreiro} R.,
  {Rodrigo} C.,  2016, \mn@doi [\mnras] {10.1093/mnras/stw200}, \href
  {https://ui.adsabs.harvard.edu/abs/2016MNRAS.457.3396P} {457, 3396}

\bibitem[\protect\citeauthoryear{{Rauch}}{{Rauch}}{2008}]{Rauch08}
{Rauch} T.,  2008, \mn@doi [\aap] {10.1051/0004-6361:200809430}, \href
  {https://ui.adsabs.harvard.edu/abs/2008A&A...481..807R} {481, 807}

\bibitem[\protect\citeauthoryear{{Rauch} \& {Deetjen}}{{Rauch} \&
  {Deetjen}}{2003}]{Rauch03}
{Rauch} T.,  {Deetjen} J.~L.,  2003, in {Hubeny} I.,  {Mihalas} D.,   {Werner}
  K.,  eds,  Astronomical Society of the Pacific Conference Series Vol. 288,
  Stellar Atmosphere Modeling. p.~103 (\mn@eprint {arXiv} {astro-ph/0403239})

\bibitem[\protect\citeauthoryear{{Riello} et~al.,}{{Riello}
  et~al.}{2020}]{Riello20}
{Riello} M.,  et~al., 2020, arXiv e-prints, \href
  {https://ui.adsabs.harvard.edu/abs/2020arXiv201201916R} {p. arXiv:2012.01916}

\bibitem[\protect\citeauthoryear{{Skrutskie} et~al.,}{{Skrutskie}
  et~al.}{2006}]{Skrutskie06}
{Skrutskie} M.~F.,  et~al., 2006, \mn@doi [\aj] {10.1086/498708}, \href
  {http://adsabs.harvard.edu/abs/2006AJ....131.1163S} {131, 1163}

\bibitem[\protect\citeauthoryear{{Taylor}}{{Taylor}}{2005}]{Taylor05}
{Taylor} M.~B.,  2005, in {Shopbell} P.,  {Britton} M.,   {Ebert} R.,  eds,
  Astronomical Society of the Pacific Conference Series Vol. 347, Astronomical
  Data Analysis Software and Systems XIV. p.~29

\bibitem[\protect\citeauthoryear{{Taylor}}{{Taylor}}{2006}]{Taylor06}
{Taylor} M.~B.,  2006, in {Gabriel} C.,  {Arviset} C.,  {Ponz} D.,   {Enrique}
  S.,  eds,  Astronomical Society of the Pacific Conference Series Vol. 351,
  Astronomical Data Analysis Software and Systems XV. p.~666

\bibitem[\protect\citeauthoryear{{Wolf} et~al.,}{{Wolf} et~al.}{2018}]{Wolf18}
{Wolf} C.,  et~al., 2018, \mn@doi [\pasa] {10.1017/pasa.2018.5}, \href
  {https://ui.adsabs.harvard.edu/abs/2018PASA...35...10W} {35, e010}

\bibitem[\protect\citeauthoryear{{Wright} et~al.,}{{Wright}
  et~al.}{2010}]{Wright10}
{Wright} E.~L.,  et~al., 2010, \mn@doi [\aj] {10.1088/0004-6256/140/6/1868},
  \href {http://adsabs.harvard.edu/abs/2010AJ....140.1868W} {140, 1868}

\makeatother
\end{thebibliography}




\appendix

\section{Virtual Observatory compliant, online catalogue}

In order to help the astronomical community on using our
catalogue of 
hot sds candidates,
we developed an archive system  that  can  be  accessed  from  a  webpage\footnote{\url{http://svocats.cab.inta-csic.es/hsa2/}} or  through  a
Virtual Observatory ConeSearch\footnote{e.g.\url{http://svocats.cab.inta-csic.es/hsa2/cs.php?RA=155.841&DEC=-37.617&SR=0.1&VERB=2}}.

The  archive  system  implements  a  very  simple  search
interface that allows queries by coordinates and radius as
well as by other parameters of interest. The user can also select the maximum number of sources (with values from 10 to
unlimited).
The result of the query is a HTML table with all the
sources found in the archive fulfilling the search criteria. The
result can also be downloaded as a VOTable or a CSV  file.
Detailed information on the output  fields can be obtained
placing the mouse over the question mark located close
to the name of the column. The archive also implements the
SAMP\footnote{\tt http://www.ivoa.net/documents/SAMP}
(Simple  Application  Messaging)  Virtual  Observatory  protocol.  SAMP  allows  Virtual  Observatory  applications  to  communicate  with  each  other  in  a  seamless  and
transparent manner for the user. This way, the results of a
query  can  be  easily  transferred  to  other  VO  applications,
such as, for instance, Topcat.



\bsp	
\label{lastpage}
\end{document}